# Seeing atoms with an atomic force microscope

Zach Cresswell and Jason Kawasaki
Materials Science and Engineering, University of Wisconsin - Madison

**Purpose**: To understand how scientists "see" individual atoms, and why this is important

**Learning Objectives:**
1. Learn how scientists can see the atomic structure of materials (such as metals, semiconductors, and ceramics), using an atomic force microscope (AFM)
2. Apply engineering principles to design and build a working model of an AFM
3. Understand that the atomic structure of a material affects its properties

**Brief description:**
In this activity, students will make a working model of an atomic force microscope (AFM). A permanent magnet attached to a compact disc (CD) strip acts as the sensor. The sensor is attached to a base made from Legos. Laser light is reflected from the CD sensor and onto a sheet of photosensitive paper. An array of permanent magnets attached to cardboard acts as the atoms on a surface. When the sensor is brought near this atomic surface the magnets will deflect the sensor, which in turn deflects the reflected laser. This deflection is recorded on the photosensitive paper, which students can take home with them.

[video demo](#)

**Next Generation Science Standards:**

**Grade level:**

**Time:** 45 minutes

**Materials:**
For a class size of 20 students (Figure 1):
- [eggcrate foam](#) (to help with the introduction)
- AFMs:
  - Tape
  - Scissors

- 3 rolls [mounting tape](#)
- [Lego blocks](#) (~350 assorted duplo blocks in total): this is used to make the base
- CDs for making the sensor. Only one sensor can be made from each CD, so have spares in case they need to make a new one
- 10x 2oz tub [play-doh](#): for mounting the laser
- Sample test surfaces
  - 1x rectangular strips of cardboard, around 10" long
  - 100 assorted-size [neodymium magnets](#)
    - Other assorted magnetic material optional (screws, weak magnets, paperclips, etc)
- Testing stations
  - 10x [5mW laser pointer](#) (blue 405nm light): used to track the deflection of the CD sensor. It must be a violet laser in order to react with the photosensitive paper. Other colors of laser will work for demonstration purposes, but in general these will not leave a trail on photosensitive paper.
    - [laser pointer batteries](#)
    - 10x zip tie: this is used to hold the laser switch
  - 5x 10x15" [Lego baseplate](#)
  - 100x [2x2 flat tiles](#):

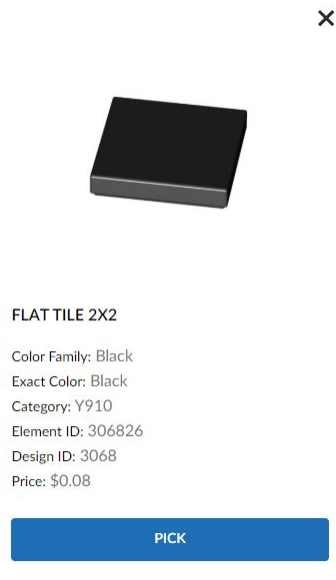

Screenshot of flat lego tile taken from LEGO website

- 5x 12-pack [photosensitive paper](#)
- 5x rectangular sheet of cardboard, around 8"x11"

- 1 [Set of lego wheels](#)

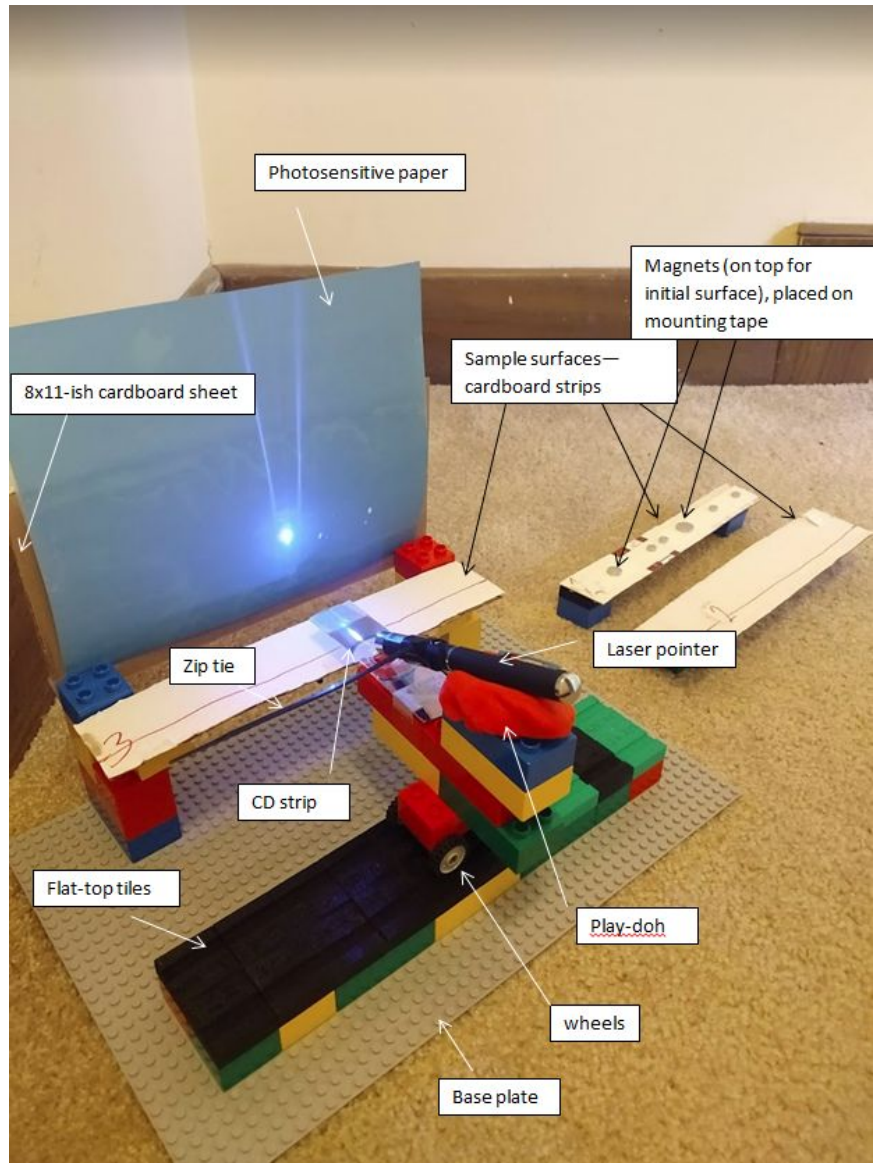

Figure 1. Materials required for this activity. This picture contains at least one copy of all the materials needed for this lesson, as well as what everything should look like when put together.

**Introduction:**

All materials, including wood, concrete, ceramics, metals, and even the human body, are made of atoms. Atoms are the fundamental building blocks of all materials. They combine together in the same way that individual Legos can be stuck together to form larger objects like cars, houses etc (Figure 2).

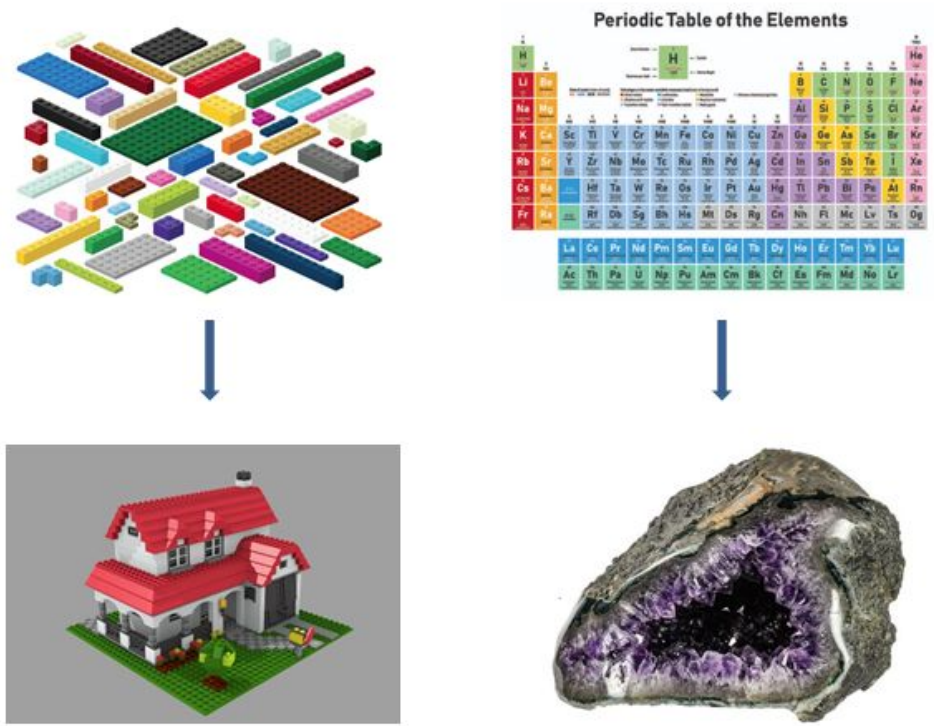

Figure 2. Atoms combine to make a material in the same way legos combine to make houses

The properties of materials--including hardness or softness, whether it conducts heat or electricity, whether it is transparent or opaque--all depend on how these atoms (Lego pieces) are arranged. A good example of this is the differences between graphite and diamond. They are both made completely out of carbon, but the difference in the patterns of their respective carbon atoms makes diamond extremely hard, insulating, and somewhat transparent while graphite is soft, conductive, and dingy gray.

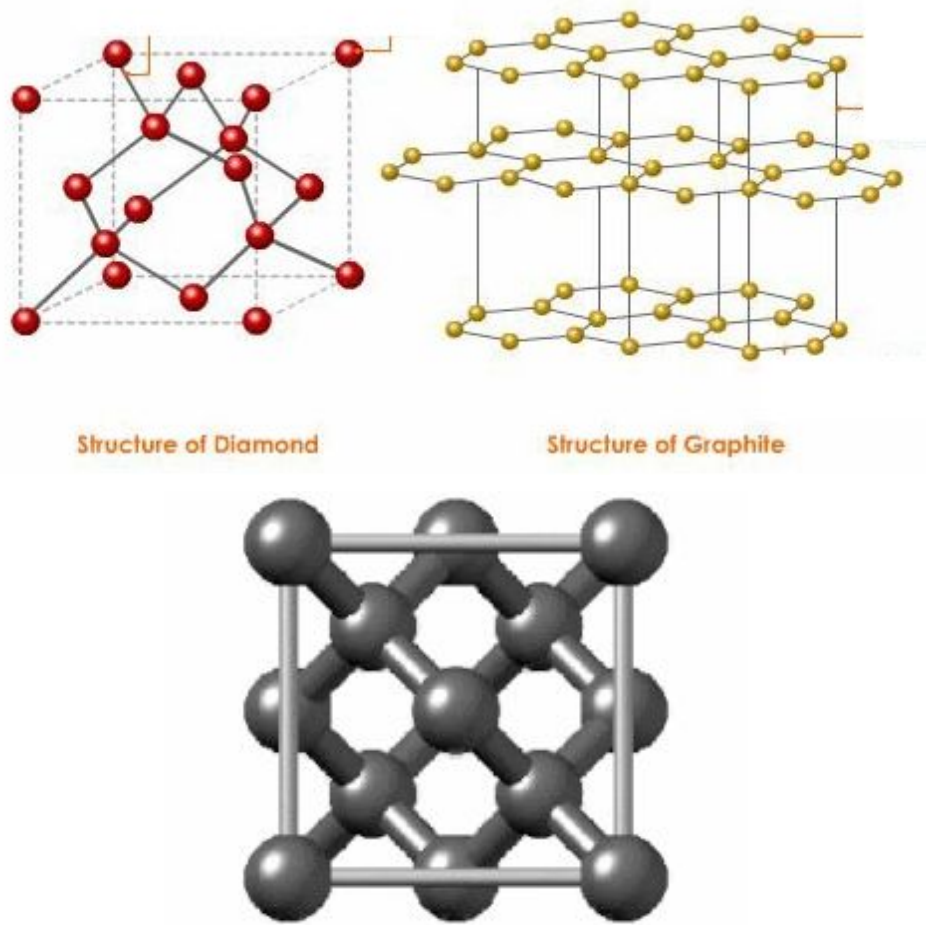

Figure 3. Atomic patterns of diamond (left) and graphite (right) [4], as well as a video of the diamond structure being rotated [5]

So, knowing this information is a very important part of studying any given material. Knowing this atomic pattern is one of the first steps to figuring out what a material is and what it can do. However, atoms are incredibly small: they are approximately one million times smaller than the width of a human hair! This is much too small to see with a normal microscope. So how can scientists tell how the atoms are arranged in a material without being able to see them?

Although we can't "see" atoms, we can "feel" their presence in the same way that one can feel a rough surface. One device that helps materials scientists feel atoms is called an atomic force microscope (AFM). An AFM consists of a narrow bar, called a cantilever, with a sharp needle attached to the end (Figure 4). This needle is dragged across the surface of a material, and when the needle comes into contact with an atom, it causes the cantilever to bend upwards. Using a laser, scientists can measure how much the cantilever bends, and that tells them where each atom is located and how "tall" each atom is.

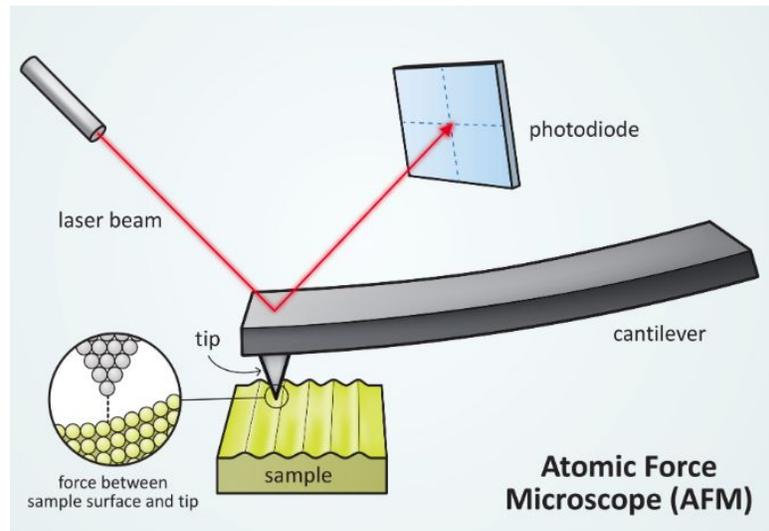

Figure 4. The main components of an atomic force microscope (AFM). The "photodetector" is the detector that records the information, the "cantilever" is the sensor [1]

The resulting image is a "height map" of the material's surface, similar to a topographic map of the hills and valleys on the Earth's surface.

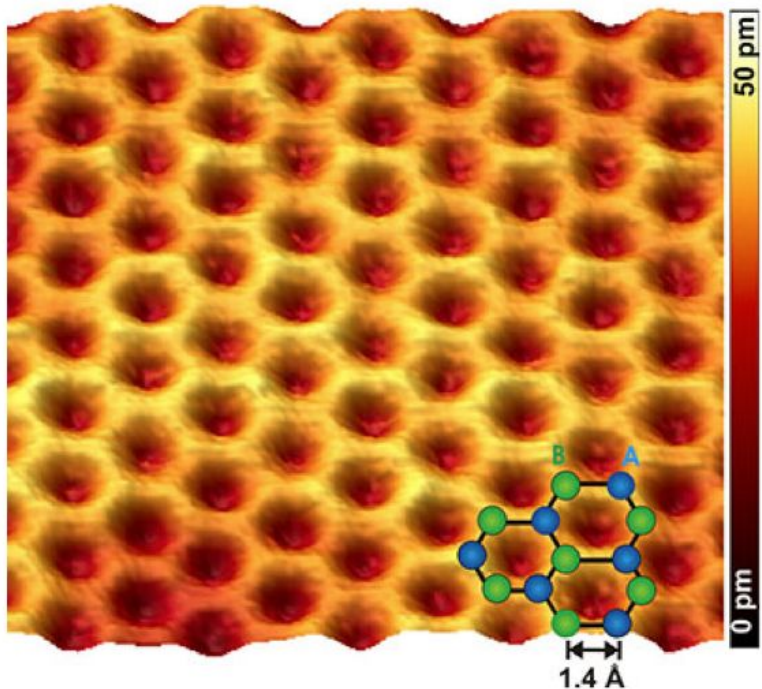

Figure 5. This image shows the kind of picture that an AFM gives. Brighter spots, where the atoms are, correspond to higher height values. [2]

In this activity students will build a working model of an AFM. Their model AFM will use magnetic force instead of mechanical (ie, a small magnet attached to the sensor instead of a small needle), but otherwise all concepts remain the same. They will test their AFMs on mock samples, with an array of magnets acting as the "atoms". The height profile data will be recorded on photosensitive paper, which students can take home with them. The activity is based on other home-built AFM demonstrations, as described in Ref 3.

**BEFORE CLASS TIME:**
1. Organize the neodymium magnets
	-Mark either the N or S side of each magnet using a marker to make it easier to use them later

2. Make the "samples"
	-Attach magnets of varying sizes, spaced randomly along a straight line, to the underside of each of the 10" cardboard strips (Figure 6). Make sure that the hidden magnets are all aligned so that they all repel the one attached to the students' AFMs later on.
	-Design it so that each "surface" has a different level of difficulty. You can do this in a few ways:

-The smaller the magnets are, the tougher it will be to find them
-The closer the magnets are to each other, the more difficult they will be to distinguish.
-Some other weaker magnetic items, like bottle caps or paper clips, can also be used to add to the difficulty (don't have to orient these in any particular way)
-Attach legos to each sample surface so they can be "clicked" into place on the testing station baseplate and taken off again easily.

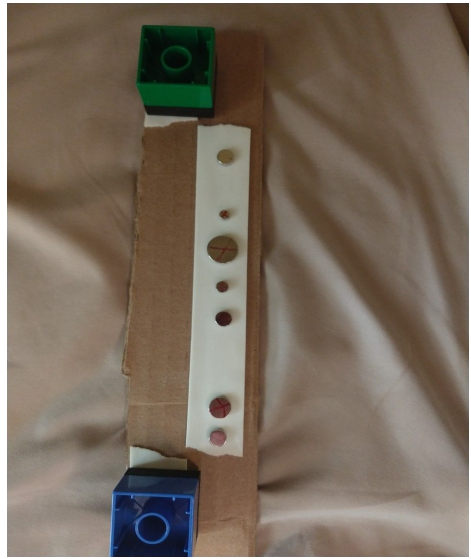

Figure 6. One possible layout of magnets hidden underneath a sample surface. This is a layout of medium difficulty

3. Set up lego analysis stations (Fig. 10)
-Lay out a 10"x15" lego baseplate
-On one end of the 15" side, place legos where the sample surfaces can be mounted
-On the other end, lay out a board of flat-top lego tiles where the AFMs can be placed and rolled around on the wheels
-Make one of these for every four students

4. Modify each laser pointer:
- Most off-the-shelf laser pointers do not have a permanent on-off switch. We find that a zip tie can be used to hold the laser in the on position.
- Close a zip tie so that it fits onto the laser pointer without falling off and position it over the on-off button. The zip tie should have a teardrop shape, so that you can rotate it to turn the laser on/off.

-Wrap tape around the side of the laser pointer closer to the lens end so that the zip tie stays in place better.
-Set up one laser pointer for every pair of students

**SHORTLY BEFORE ACTIVITY:**
1. Place CDs in hot water
    -This will make them easier to cut into strips later when the students are building their AFM
2. Build example AFM

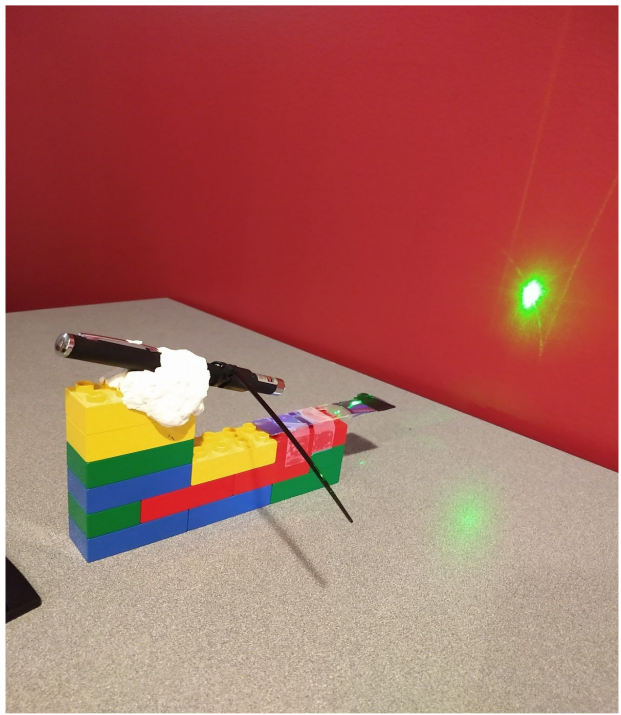

Figure 7. Another simple example AFM build, slightly different than the one above. The students should be able to create this in whatever shape they want as long as it is within size constraints. The laser is fixed to the Legos using Play-Doh.

**Procedure:**
1. Preliminary discussion
    -Pose the question: Actual atoms are much too small to be seen by our eyes, even with a microscope. How can we "see" the atoms without seeing them?
    -Introduce the idea of "feeling" atoms instead of seeing them
        -Introduce this concept as the concept behind atomic force microscopy, or AFM

-Demonstrate this by passing out acoustic foam and having the students run the tip of a pencil or pen across it, showcasing the deflection of the tip (Fig. 8). Draw parallel to the setup of an actual AFM.

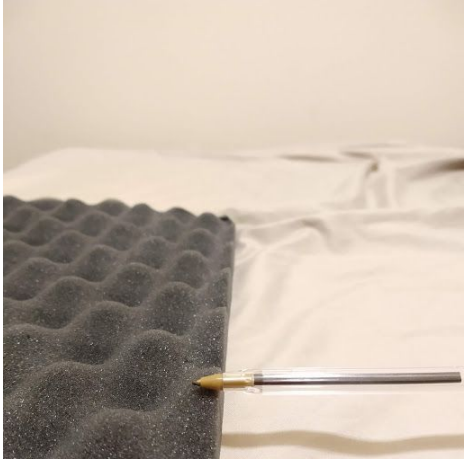

Figure 8. Egg Crate foam is a good analogy for the surface of a material

-Introduce magnetism as another "force" that can be used in the same way.
-Pair the students up, tell them that they will be building their own magnetic AFM and studying the "sample surfaces" set up earlier
-Describe the basic setup that their AFMs will need
    -Body made of legos
    -Magnet on one end attached to a reflective CD surface
    -Area on the other end to mount the laser pointer
-Show an example of a simple AFM made this way

2. Build AFMs
    -Allow the students time to come up with their AFM designs, give them an extra set of hands if they need it.
    -Give them scissors to cut the CDs into strips with (Fig. 9).

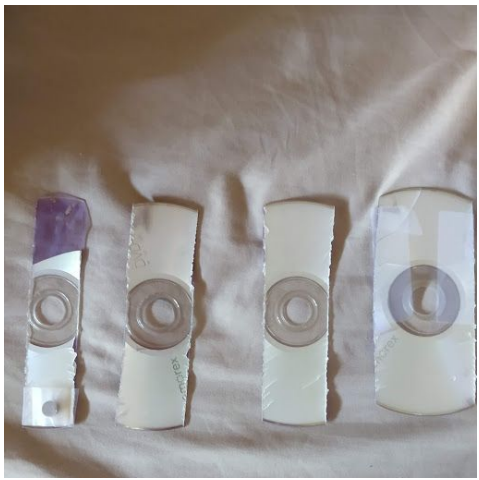

Figure 9. The cut out stips should look something like this. The cut can be any thickness. They don't even have to cut it in the middle if they don't want; let them figure out what design they think is best.

-Make sure that they orient their magnets correctly, theirs should repel the ones in the sample surfaces. This is where marking the north and south poles of each magnet comes in handy.

3. AFM magnetism test:
   -Pair groups of two up, assign two groups of two--a total of four students--to each testing station. Give each testing station three sample surfaces to analyze.
   -When the students feel their AFM is ready, have them bring it to their testing station. Connect the wheels to it and place it on the track. Click a sample surface in place on the other side of the baseplate.

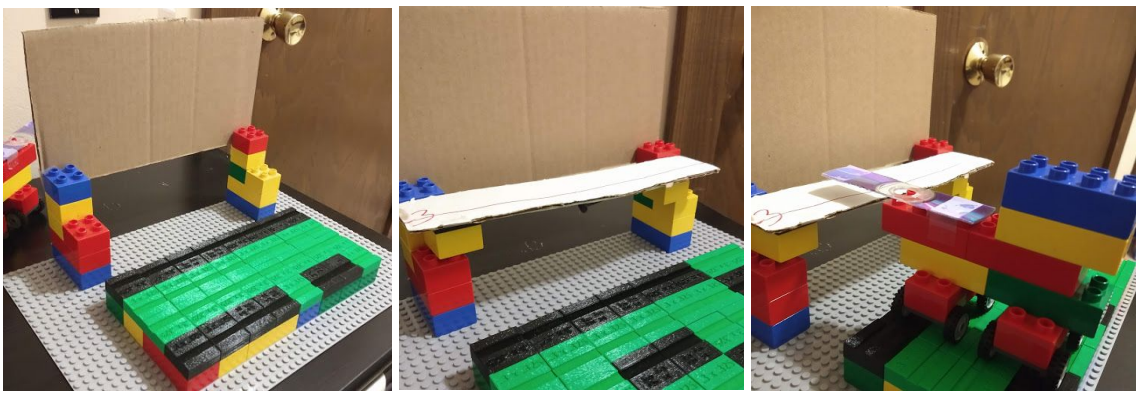

Figure 10. AFM analysis station. The sample surfaces can be clicked into place in the lower step of the side with the cardboard sheet. Make sure it is possible for the CD strip to be only slightly above the sample surface if the students build their AFMs well.

-Mount a piece of photosensitive paper on the setup, where the laser will be shining.
-Attach the wheels to their design--as many or as few as required--and set it on the flat-top tiles.
-Have them move their AFM slowly and steadily over the sample surface so that their CD strip is in line with the hidden magnets. If done correctly, the laser pointer should leave a white trail across the photosensitive paper as the AFM moves, as in the example video.
-If their setup doesn't work (if they can't sense the magnets), allow them to go back and modify their AFM and test it again later.
-If both pairs of two are successful with the premade surfaces, allow each of them to create their own surface for the group they are paired with to test
    -Give them various magnets as well as other metallic objects, tell them to arrange them as they see fit.

**Discussion:**
Talk about the students' AFM designs, and what they found worked and did not work. Talk about the best combination of the important parameters, including: the size of the magnet used in the AFM, the distance between their magnet and the magnets in the sample surface, the thickness of their CD strip, and the total length of the CD strip that was suspended in the air. Discrepancies in design like this is the heart of what separates an engineer from a regular scientist.

As a thought exercise, ask why they think a reflected laser is used to record the tip's position in a typical AFM system instead of something simpler, like maybe a pencil attached to the CD strip or something.

As to why this information is important, give the example of the difference between graphite and diamond. Hand the models of their structures out for the students to see--or just show a picture of each model from google--and discuss how the different shapes give very different properties. Have them draw what they think AFM measurements of each of them would look like if they were to use their AFMs on these sorts of patterns and have them compare the two.

**Evaluation:**
Describe how atomic force microscopes work
What parameter determines how small an object an AFM is able to detect?
Why, in your own words, is this measurement important?

**Acknowledgments**

We thank Andrew Greenberg and Matthew Stilwell for their feedback on this activity. This work was supported by the National Science Foundation (DMR-1752797 and DMR-1720415).